\documentclass[twocolumn]{bmcart}

\usepackage[utf8]{inputenc} 
\usepackage{graphicx}
\usepackage[utf8]{inputenc}
\usepackage[T1]{fontenc}
\usepackage{textcomp}
\usepackage{siunitx}

\startlocaldefs
\endlocaldefs

\begin{document}

\begin{frontmatter}

\begin{fmbox}
\dochead{Research}


\title{Wearable vibrotactile stimulation for upper extremity rehabilitation in chronic stroke: clinical feasibility trial using the VTS Glove}  

\author[
   addressref={aff1},            
   corref={aff1},                   
   email={cseim@stanford.edu}   
]{\inits{CES}\fnm{Caitlyn E.} \snm{Seim}}
\author[
   addressref={aff2}, 
]{\inits{SLW}\fnm{Steven L.} \snm{Wolf}} 
\author[
   addressref={aff3}, 
]{\inits{TES}\fnm{Thad E.} \snm{Starner}} 


\address[id=aff1]{ 
  \orgname{Stanford University},  
  \street{Department of Mechanical Engineering},
  \city{Stanford, CA},   
  \cny{USA}   
} 

\address[id=aff2]{ 
  \orgname{Emory University School of Medicine},  
  \street{Department of Rehabilitation Medicine},
  \city{Atlanta, GA},       
  \cny{USA}                
}

\address[id=aff3]{ 
  \orgname{Georgia Institute of Technology}, 
  \street{College of Computing},
  \city{Atlanta, CA},        
  \cny{USA}              
}


\begin{artnotes} 
\end{artnotes}

\newcommand\ces[1]{\textcolor{green}{#1}}



\begin{abstractbox}

\begin{abstract} 
\textbf{Objective:} 
Evaluate the feasibility and potential impacts on hand function using a wearable stimulation device (the VTS Glove) which provides mechanical, vibratory input to the affected limb of chronic stroke survivors.

\noindent\textbf{Methods:} 
A double-blind, randomized, controlled feasibility study including sixteen chronic stroke survivors (mean age: 54; 1-13 years post-stroke) with diminished movement and tactile perception in their affected hand. Participants were given a wearable device to take home and asked to wear it for three hours daily over eight weeks.  The device intervention was either (1) the VTS Glove, which provided vibrotactile stimulation to the hand, or (2) an identical glove with vibration disabled.  Participants were equally randomly assigned to each condition. Hand and arm function were measured weekly at home and in local physical therapy clinics.

\noindent\textbf{Results:}  
Participants using the VTS Glove showed significantly improved Semmes-Weinstein monofilament exam, reduction in Modified Ashworth measures in the fingers, and some increased voluntary finger flexion, elbow and shoulder range of motion.  

\noindent\textbf{Conclusions:}  
Vibrotactile stimulation applied to the disabled limb may impact tactile perception, tone and spasticity, and voluntary range of motion.  Wearable devices allow extended application and study of stimulation methods outside of a clinical setting.

\end{abstract}


 
\end{abstractbox}

\end{fmbox}

\end{frontmatter}


\section*{Background}
Over 15 million people have a stroke each year, making it one of the leading causes of disability in the United States and worldwide \cite{bonita2004global, centers2012prevalence, adamson2004stroke}. 
Upper limb disability occurs in about 50\% of cases \cite{wade1983hemiplegic, parker1986loss} and diminished tactile perception in about 35-55\% \cite{connell2008somatosensory, tyson2008sensory}.  Current methods of therapy for upper limb dysfunction after stroke focus on activities which use the limb; however, these forms of rehabilitation are not accessible to survivors with very limited function.

Somatosensory stimulation may be an effective and accessible modality for rehabilitation.  
Most fundamentally, somatosensory input is known to drive cortical organization and skill acquisition \cite{feldman2005map, van1973somatosensory, buonomano1998cortical}.   
Somatosensory input has also been associated with sensorimotor recovery after CNS injury in animal  \cite{xerri1998plasticity, jablonka2010remapping} as well as human studies \cite{bird2013sensorimotor}. 
Afferent input is also integral to limb use.   
Tactile perception and proprioception are factors in motor performance and are thought to co-activate with motorcortical circuits \cite{pleger2003pharmacological, porter1992patterns, mattay1998hemispheric}.

\begin{figure}
    \centering
    \includegraphics[width=.75\columnwidth]{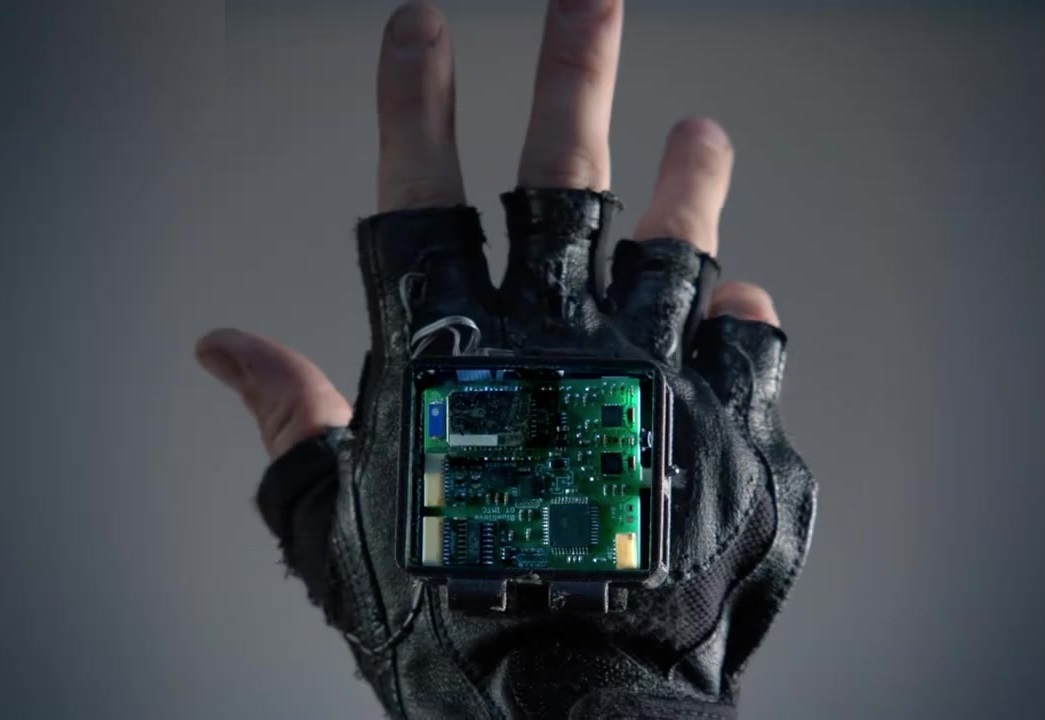}
    \caption{The computerized glove that provides vibrotactile stimulation for this study.}  
    \label{g0}
\end{figure}

Afferent electrical stimulation has been studied as a means for providing sensory input to the disabled extremity of stroke survivors \cite{dewald1996long, schabrun2009evidence, dimitrijevic1996modification, smith2009effects, tu2017effects}, and preliminary evidence shows changes in tactile perception, motor function and brain activity.  Mechanical, vibratory stimulation can be applied without the placement, electrodes or gel of electrical stimulation.  Afferent electrical stimulation most often targets cutaneous sensory receptors in the skin; while vibrotactile stimulation can activate both muscle afferent fibers and cutaneous sensory receptors without inducing movement.  Vibrotactile stimulation has been coupled with other methods such as robotic manipulation or music practice exercises for rehabilitation \cite{cordo2009assisted, estes2015wearable}, and applied to the arm for in-situ dexterity improvement \cite{enders2013remote}.  
Improved spasticity and significant neuromuscular changes have been found in laboratory studies of whole-body vibration (WBV) \cite{abercromby2007variation, delecluse2003strength, ness2009effect, pang2013effects} and focal muscle/tendon vibration \cite{binder2009vibration, kossev2001crossed, marconi2011long, noma2012anti, rittweger2010vibration, siggelkow1999modulation,steyvers2003frequency}.   

Despite encouraging data, vibrotactile stimulation is not widely used outside the clinic because there are no mobile devices that can deliver and study this form of mechanical stimulation for prolonged periods of time.  
Here we designed a lightweight, wireless, wearable device to apply vibrotactile stimulation to the hand.  Wearable devices are closely coupled with the body, and thus allow stimulation for extended periods of time and in the background of daily life.    The intervention is mobile and simple to apply without access to a clinic.  Users simply wear the device, requiring little exertion and time, which may facilitate adherence. 
The device was deployed in a controlled feasibility trial of chronic stroke survivors with upper limb sensorimotor deficits.  If wearable stimulation proves to be effective it could directly impact healthcare delivery, because it may provide a mobile, affordable rehabilitation option for patients who otherwise would not have access to high intensity stroke rehabilitation.

\section*{Methods}
The study was a double-blind, randomized controlled study performed
in Atlanta, Georgia.
Eligible participants were randomly assigned to the vibrotactile stimulation glove (VTS) or sham control glove (control) condition.  All were asked to wear the device on their affected hand for three hours each day for eight weeks.  
As a feasibility study, the trial was not listed with clinicaltrials.gov but was approved and overseen by the Office of Research Integrity's IRB board of Georgia Institute of Technology.    
All participants were screened using the Mini Mental State Exam (MMSE) and provided written consent before beginning the study.

\subsection*{Participants}
\begin{center}
    
\begin{figure*}
\centering
\includegraphics[width=1.99\columnwidth]{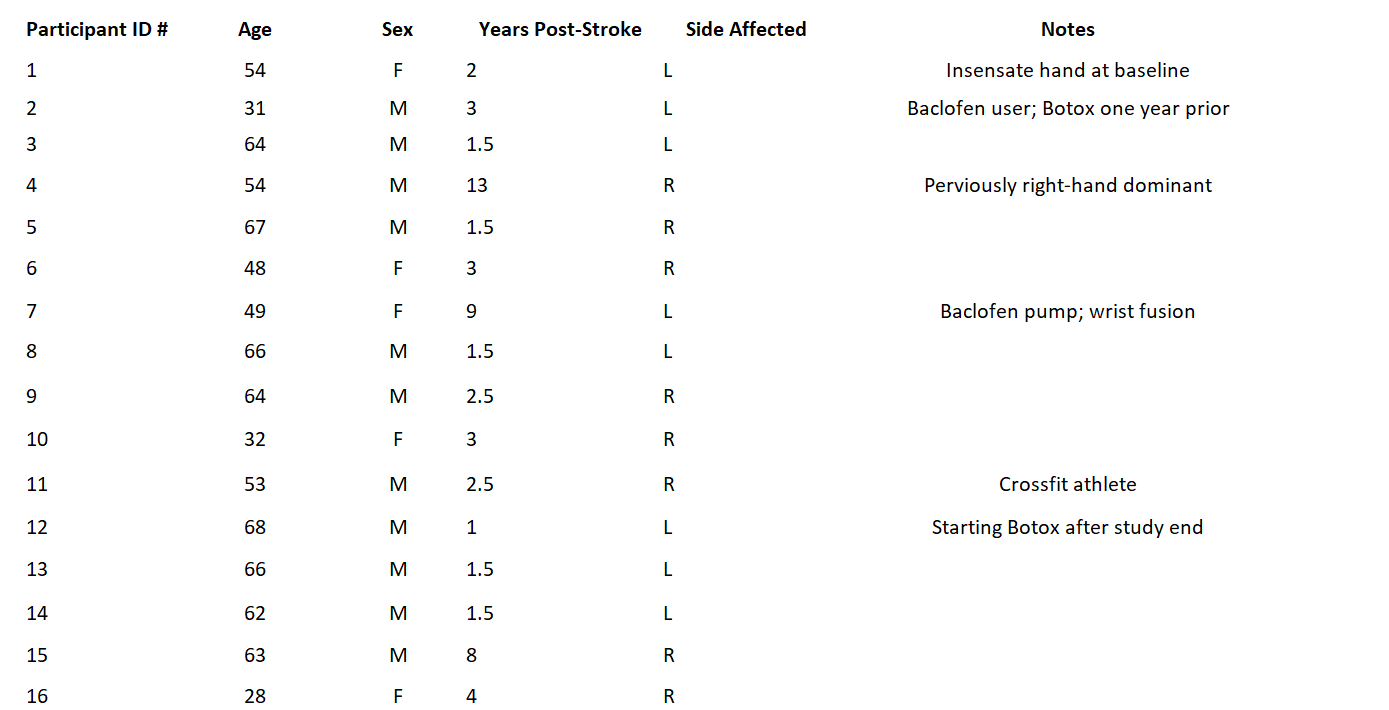}
\caption{Demographics and notes for participants in the study.  The experimental VTS group includes participants 1-8, and the sham control group includes participants 9-16.  These participant numbers were assigned only to present data in this manuscript.}
\label{demog}
\end{figure*}
\end{center}

The study included 16 chronic stroke survivors with upper extremity deficits
(ages 28-68; 1-13 years post stroke (Mean=3.7, SD=3.3); 8 VTS condition/8 control condition). 
Participants were recruited through stroke support groups in the Atlanta metropolitan area.  Figure \ref{demog} shows a breakdown of participant demographics. Individuals with various levels of arm function could participate.  The protocol requires no exercises and thus is accessible to patients with very limited movement.  Because this investigation is preliminary, no prior data are available for optimal sample size calculation. 

\subsubsection*{
Inclusion criteria:}
\begin{itemize}
    \item History of stroke \textgreater 1 year prior
    \item Impaired touch sensation in the hand (Semmes-Weinstein monofilament exam score of $\geq$ 0.2 grams on 3 of 20 measured locations on the hand)  
    \item Passive range of motion allows user to don a glove
    \item English speaker, age 18+ 
\end{itemize}

\subsubsection*{
Exclusion criteria:}
\begin{itemize}
    \item Intact sensation in the hand (determined by Semmes-Weinstein monofilament exam)
    \item Active Range of Motion within normal limits for all joints of the fingers
    \item Cognitive deficits, dementia or aphasia (MMSE score of \textless22) that prevent informed consent
    \item Other neurological condition that may affect motor response (e.g. Parkinson’s, ALS, MS)
    \item Pain in the limb that substantially interferes with ADLs or prior arm injury
    \item Enrollment in a conflicting study, Botox treatment, or other upper extremity rehabilitation program during the study period
\end{itemize}

\subsection*{Study Design}
The study consisted of eight weeks using the stimulation or sham device during daily life. Participants wore the glove daily and met with blinded study administrators for weekly visits to measure sensorimotor function.   

At the first visit, all participants received a device, cord and safety manual to take with them.  Participants were instructed to \textbf{wear the device, turned on, for three hours every day} while awake.  Users were notified that an onboard measurement unit would track usage time, and that 21 hours of weekly use is required.  
All participants were advised to charge the glove each night using the cord provided, just as one might do with a cell phone. Then, wear the device on-the-go or at home during their normal routine.  Wearing did not need to be continuous each day, but had to total three hours.  The dosage was chosen to be intensive, while not too much daily commitment for participants. 

\subsection*{Apparatus}
A wearable computing glove was designed to provide vibrotactile stimulation to participants throughout their daily life (Figure \ref{g0}).   It can be taken home and used outside of the clinic environment. Additionally, the glove is worn while users conduct their daily life – making the rehabilitation low-effort and ``passive.''  

\subsubsection*{Wearable Device}
The wearable device (``VTS Glove’’) is designed to be low-cost, lightweight, and mobile.  The device is a fingerless glove with a vibration motor attached to each dorsal phalanx.  This design allows a designated actuator for each finger, while stimulating a region where vibrations can reach the glabrous skin of the palm and the finger extensor tendons.  
The heart of the device is a circuit board and microcontroller, which activates these motors in a pre-programmed sequence when the switch is turned ``on.’’  The onboard gyroscope logs movement data along with usage data onto a microSD card which is checked by proctors for protocol adherence each week. The glove is rechargeable and has a battery life that allows wireless stimulation for four hours between charges.  Design and implementation of the device is reported in detail in a companion manuscript \cite{seim2019wearable}.

\subsubsection*{Stimulus Design}
For this experiment, stimulation characteristics were designed to target cutaneous mechanoreceptors -- specifically the Pacinian corpuscles -- which respond to direct vibration and vibration transmitted through the body at a frequency range of 10-400 Hz (preferentially responding around 250 Hz) \cite{johnson2001roles}.  Stimulation pattern and timing was designed to be intensive but not uncomfortable by using many vibration pulses with a changing location across the fingers.

Small, coin-shaped vibration motors from Precision Microdrives (ERM-type, Model \#310-113) provide the stimulation for this experiment.  These motors were driven at a voltage of 3.3V for an approximate amplitude of 1.5 g and 210 Hz vibration frequency (measured in a laboratory setting for validation at 1.3 g and 175 Hz when attached to the glove).  Two stimulation sequences were used, each based on the finger pattern for a piano song.  Song patterns provided a framework for pseudo-random stimulation and the option to later combine stimulation with music practice exercises for a lighthearted therapy routine.  Each song pattern (Ode to Joy and Happy Birthday) was extended with a short sequence to balance stimulation evenly across all fingers.   
During each repetition the pattern played once quickly (250 ms vibrations, 100 ms pause between each stimulus) and once slowly (700 ms vibrations, 100 ms pauses).  These songs were chosen for their recognizable, one-handed melodies with 5-7 notes which could be played on the keyboard with little-to-no hand shifting. The stimulation pattern was switched weekly, alternating between the two ``songs.’’

 \subsection*{Conditions}
Participants continued their standard of care, and none were enrolled in concurrent upper limb rehabilitation programs.  

\subsubsection*{Intervention Condition}
Participants in the vibrotactile stimulation (VTS) condition received a glove with vibration enabled.   The protocol used here includes no required exercises.   Participants were asked to wear their glove, switched on (so the indicator light appears), for three hours daily while awake.  Users should also charge the battery each night and as needed.

\subsubsection*{Control Condition} 
Participants in the sham control condition receive a glove with vibration disabled. 
The appearance of the device was the same as the experimental condition. 
All indicator lights on the computer board activate in the same fashion.   
Instructions and language also matched those in the VTS condition: wear the glove on their affected hand, switched on, for three hours daily while awake, and charge the battery each night.  

The control condition was assigned a sham device (rather than no intervention) to examine the tolerance of the wearable device with and without stimulation, evaluate if the vibrotactile stimulation itself may have an impact on measures, and provide some data on mechanisms underlying this technique by comparing the conditions.    

\subsection*{Outcome measures}
Baseline demographic information collected was sex, age, date of stroke, type of stroke, and side affected.  Measurements are taken during weekly visits throughout the study.  Visits occur at the patient's home or a midway meeting spot.  
All measures were performed by trained proctors not involved in the intervention or data analysis.  For all participants, key measurements were taken by a blinded occupational therapist.  Those measures were taken at the beginning (day 0), middle (4 weeks), and end (8 weeks) of the study.  The therapist and study proctor for each participant was consistent to minimize inter-rater variability. 

The intent of this study was to examine the initial feasibility in this device and technique. Thus, data on engineering, design, comfort and usability was collected through weekly surveys and observations.  
Engineering data are presented in another manuscript along with subsequent design work \cite{seim2019wearable}.  While here we provide data on measures of arm function. 

\subsubsection*{Primary Outcome Measures}
The Semmes-Weinstein Monofilament Exam (SWME) is used to assess cutaneous tactile perception in the affected hand.  Locations on the dorsal and volar side of the hand are assessed, including the fingers.  This test has good intra-rater reliability and requires little training.  This assessment was done weekly.

\subsubsection*{Secondary Outcome Measures}
The Modified Ashworth Scale (MAS) is used to assess resistance to passive motion from involuntary muscle tone and spasticity \footnote{hypertonicity, spasticity, and resulting involuntary muscle tone}.  In this study, MAS was measured for flexion and extension of the fingers, thumb, wrist, elbow and shoulder of the participant's affected upper limb.   Confounding factors for this measure were controlled whenever possible including: time of day, time after medication dosage, arm position, and rater. 
Voluntary angular range of motion (Active Range of Motion (AROM)) is used to assess motor impairment.  Here, these measures were made for flexion and extension of the fingers, wrist, elbow and shoulder of the participant's affected upper limb.  This measure can capture changes in function when participant dexterity is too low to perform tests like the Jebsen-Taylor.  A trained occupational therapist performed all movement and spasticity measures in a clinical setting at the beginning, middle and end of the study.
Each week, participants are also given a worksheet to report what they did while wearing the device, observations, or comments about the device.

\begin{figure}
    \centering
    \includegraphics[width=.99\columnwidth]{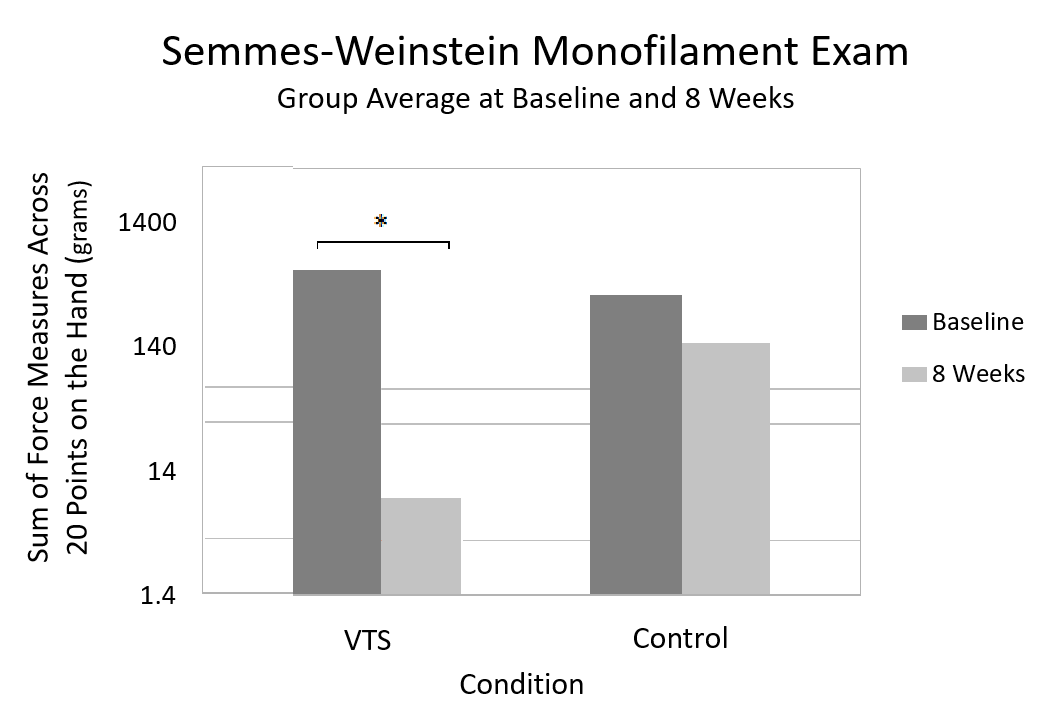}
    \caption{Semmes-Weinstein Monofilament Exam results by group at baseline and eight weeks.  This graph shows the group's average sum of perceived forces across 20 locations on the hand. Smaller perceived forces equate to greater tactile perception. Logarithmic scale used to render all force levels.}
    \label{fig:swmemeans}
\end{figure}

\begin{figure}
    \centering
    \includegraphics[width=.94\columnwidth]{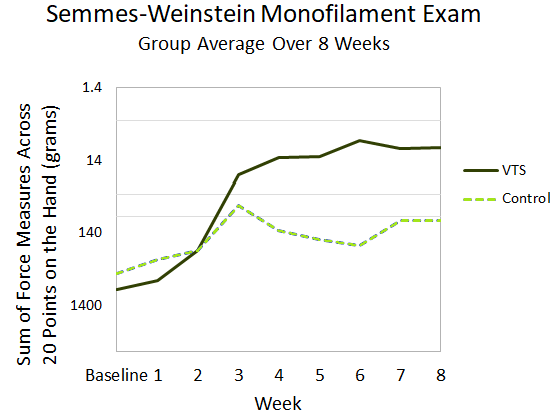}
    \caption{Trajectory of Semmes-Weinstein Monofilament Exam results over eight weeks for both conditions.  This graph shows the group's average sum of perceived forces across 20 locations on the hand.  Smaller perceived force values equate to greater tactile perception. Logarithmic scale used to render all force levels.}
    \label{fig:swmetrends}
\end{figure}


\subsection*{Data analysis}
Using an intention-to-treat analysis, we processed data for all participants including two who had to withdraw prematurely due to unrelated circumstances.
The last measured values were used for the determination of any missing values in the case of dropouts or a missed visit, conservatively assuming that no changes occurred since the last measure.  Paired observations were compared using the Wilcoxon signed-ranks test, and measures between groups were compared using the Mann–Whitney U test.  A p-value \textless 0.05 was considered statistically significant.  

\section*{Results}
Adherence for users in both conditions was measured each week using self-reported usage times matched with data from the glove's inertial measurement unit.  If usage time had not been within three hours of the required weekly time (21 hours) for two consecutive weeks, the participant would have been released from the study.  No such occurrences happened during the trial.  

\subsection*{Semmes-Weinstein Monofilament Exam (SWME)} This measure was taken at 20 points on the hand -- yielding one minimum perceivable force value per location. 
Each filament represents a level of force, but the Semmes-Weinstein exam filaments are not equidistant (i.e. 2 grams, 4 grams, and 400 grams are sequential ratings).  For this reason, a weighted sum or weighted average is calculated, to compare the sensitivity of the hand as a whole (20 points measured).  
A minimum sum of 1.4 grams corresponding to ``normal'' sensation at all points and a maximum sum of 6000 grams corresponding to only ``deep pressure sensation'' at all points. These sums were divided by 20 to calculate the average perceivable force across the hand.  Smaller perceived forces equate to better tactile perception.  
One participant in the VTS condition is not included in these calculations because their starting measures prevent representation on the graphs.  This user initially presented as insensate at all points, but could accurately report deep pressure sensation at three points later in the study.  
 
Starting means\footnote{Mean sum of force levels for the hand (sum of 20 points).} (M=832.4 grams, SD=1206 for VTS; M=501.6 grams, SD=949.7 for control) were compared using Mann–Whitney U test (U=18; z=-1.10; p=0.271). 
Baseline measures of the VTS experimental group were compared to measures at eight weeks (M=9.701 grams, SD=14.25) and results suggest that there is a significant difference (t-test: t(6)=-3.50; p=0.006; signed-ranks: Z=-1.89; p\textless0.05).  
As figure \ref{fig:swmemeans} shows, the VTS condition is able to sense smaller forces than the control condition at eight weeks (M=91.15  grams, SD=224.1).  The sham control condition also showed a change in SWME measures, but this change was not statistically significant (t-test: t(7)=1.190; p=0.254; signed-ranks: Z=-1.40; p\textgreater0.05).  
Figure \ref{fig:swmetrends} shows the trends in these values throughout the entire study.

\begin{figure*}
    \centering
    \includegraphics[width=1.9\columnwidth]{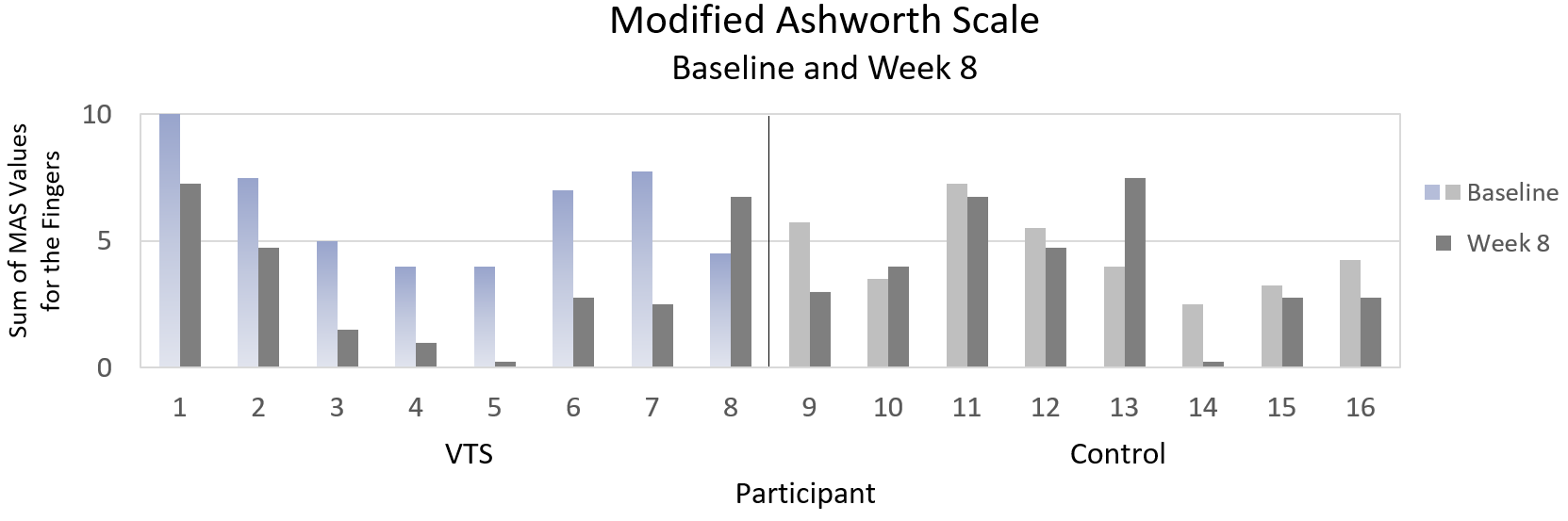}
    \caption{
Sum of Modified Ashworth values for the fingers at baseline and after eight weeks. MAS values here are reported on a scale of 0-5.  Lower scores are better.}
    \label{fig:masall}
\end{figure*}


\subsection*{Modified Ashworth Scale (MAS)}
Modified Ashworth Scale (MAS) was measured in a clinical setting for flexion and extension of MCP/PIP finger joints, thumb, wrist, elbow, and shoulder.  Here, results are reported for the fingers which showed the most change in values.  MAS values here are reported on a scale of 0-5 and summed for the fingers.   
Starting means (M=6.28, SD=2.16 for VTS; M=4.5, SD=1.46 for control) were compared using a Mann-Whitney U test (U=16; z=1.63; p=0.052) 
and no significant difference was found.   All users' starting sums can be found in Figure \ref{fig:masall}.
Differences in experimental group MAS were found to be significant
using a Wilcoxon signed-ranks test comparing starting measures to measures at 
8 weeks (Z=-2.38; p\textless0.05).  Average difference at 8 weeks was M=-2.94 total points on the Ashworth scale for the finger joints of the affected arm for users in the VTS condition.
Differences in control group MAS at 8 weeks were also compared using the Wilcoxon signed-ranks test (Z=-0.51; p\textgreater0.05)  
but the difference (M=-0.53) was not considered significant. Change from baseline was compared between conditions and found to be significantly different (Mann-Whitney: U=7.5; z=2.52; p=0.006
).  
Participant 7 had severe spasticity before the study, which led to a Baclofen pump and wrist fusion surgery. These interventions were failing to stop the progression of tone and spasticity in their hand; however, their tone was reduced after participation in the study.
Two users (5 and 7) agreed to follow-up six months post-study. There was no significant relapse in values at follow-up vs. study end.

\begin{figure*}
    \centering
    \includegraphics[width=1.99\columnwidth]{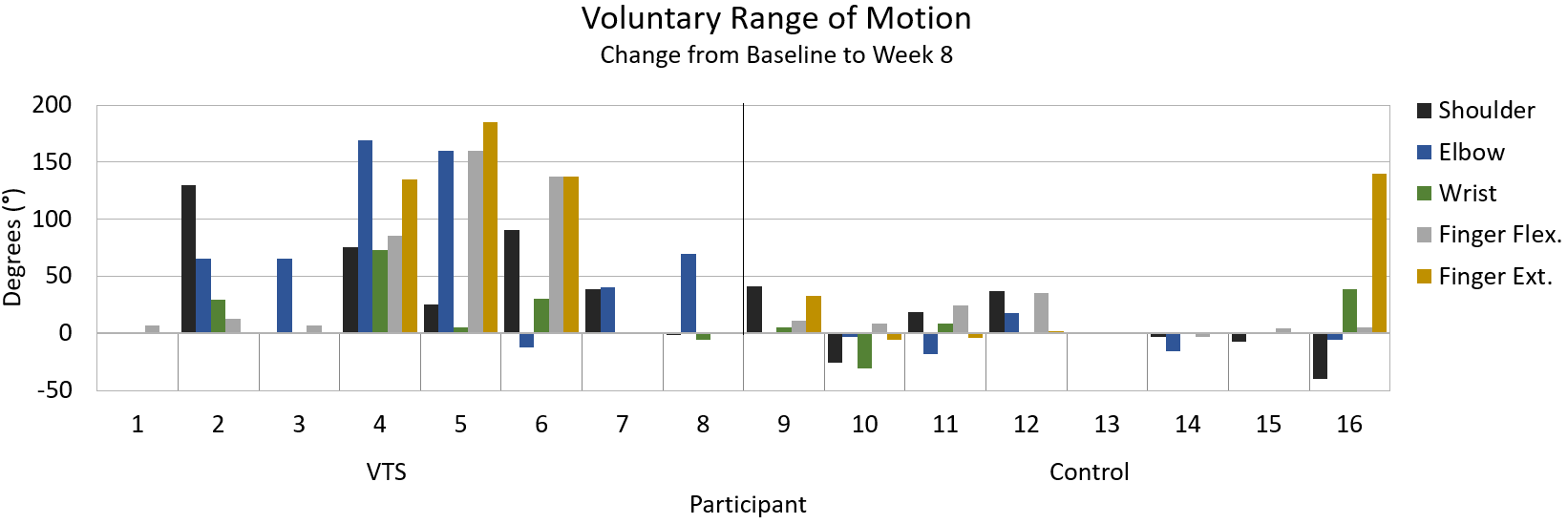}
    \caption{Increase in angular degrees of voluntary movement for four upper body locations between baseline and study end.   Shoulder, elbow and wrist values include both flexion and extension (from flexed) ranges. Finger flexion and extension is shown separately to provide greater detail, and these values include both MCP and PIP ranges. Zero values most often occurred when the participant had no voluntary movement in the joint at baseline and eight weeks.}
    \label{fig:arom1}
\end{figure*}

\subsection*{Active Range of Motion (AROM)}
Motor function was measured in a clinical setting as the angular degrees of voluntary movement at joints in the fingers and arm.  Each of four body areas are summed (i.e., voluntary angular motion for shoulder is the sum of flexion, extension and abduction).  Measures are taken in the neutral gravity plane whenever possible. Compensation from other muscles and synergy with spasticity are not included as voluntary range.  
\textbf{Finger and elbow extension is measured from a flexed position, not from neutral}, so as to report voluntary extension that may be used for activities such as releasing objects from grasp.
Finger AROM is measured at the MCP and PIP joints, and those values are summed. Thus, ``Finger Flex.'' and ``Finger Ext.'' include change in both the average MCP and PIP ranges. 

Starting means for arm motion and finger flexion had a significant difference between conditions.  The control group included fewer members with low to moderate starting function.  Baseline function may be a factor in the results for the control group, but further study is needed to examine its influence.  The control group showed no significant difference in shoulder (M=202.9\si{\si{\degree}}, SD=135.0\si{\degree}, Avg.Change=2.7\si{\degree}), elbow (M=113.8\si{\degree}, SD=115.7\si{\degree}, Avg.Change=-2.9\si{\degree}), wrist (M=48.6\si{\degree}, SD=38.9\si{\degree}, Avg.Change=3.0\si{\degree}), finger flexion (M=25.0\si{\degree}, SD=31.3\si{\degree}, Avg.Change=50.9\si{\degree})  or finger extension range (M=10.9\si{\degree}, SD=18.1\si{\degree},  Avg.Change=57.2\si{\degree}). 
 
The experimental VTS condition showed improvements in sum of shoulder (M=63.5\si{\degree}, SD=66.7\si{\degree}, Avg. Change=44.6\si{\degree}), elbow (M=54.1\si{\degree}, SD=52.5\si{\degree}, Avg.Change =69.5\si{\degree}), and sum of finger flexion (M=117.8\si{\degree}, SD=71.9\si{\degree}, Avg. Change=10.6\si{\degree}) range of motion. 
A Wilcoxon signed-ranks test found these changes to be significant (Z={-2.20, -2.03, -2.20}, p\textless0.05), and this finding was consistent with a paired t-test  (t(7)={2.59, 2.98, 2.16}, p={0.018, 0.010, 0.033}).  
Change in range of motion for the wrist (M=9.5\si{\degree}, SD=12.5\si{\degree}, Avg. Change=16.5\si{\degree}) and finger extension (M=46.3\si{\degree}, SD=52.0\si{\degree}, Avg.Change=20.7\si{\degree}) was not found to be statistically significant.  Changes in voluntary ROM are shown in Figure \ref{fig:arom1}.


 \section*{Discussion}
Participants who received vibrotactile stimulation showed significant change in measures whereas those in the control group did not.  The wearable devices successfully delivered mobile stimulation throughout the duration of the study, and all participants were able to adhere to the daily wearing protocol.  

Changes in SWME measures suggest that participants showed improved tactile perception.  Figure \ref{fig:swmetrends} suggests that the trend in improvement was gradual.  
One participant reported the return of protective sensation in cases of joint hyper-extension, and one reported being able to feel the vibrations when they could not initially.

Some participants in the VTS condition provided observations that the affected hand was more open and flexible.  These observations were consistent with changes in Modified Ashworth Scale measures.   
Figure \ref{fig:masall} shows each person’s starting and ending measures.  All but one person in the experimental condition showed a reduction in MAS values.  
Participants in both conditions must frequently stretch open their affected hand to don the device, and stretching may be associated with changes in MAS. However, participants in the control group (who also stretched to don the device daily) did not show a significant change in MAS values, which suggests that stimulation rather than stretching is associated with these changes. 
Tone and spasticity lack effective or lasting treatment options, yet 40-50\% of stroke survivors with upper extremity disability may be affected \cite{dajpratham2009prevalence, watkins2002prevalence}
.  More study is needed, but this promising preliminary evidence along with that in prior work suggests that afferent stimulation may be used to address spasticity and tone.   The VTS Glove allows extended stimulation and further study of this technique.   Future work can adjust stimulation characteristics to target different sensory receptors and examine optimal settings.  
Some participants with flexed fingers struggled to don the glove device, so the design was subsequently revised for accessibility.   

Changes in voluntary range of motion may be due in part to reduction in involuntary tone.  
Some participants showed large increases in range, with near-normal finger extension and flexion at week eight. Others showed no change in voluntary range of motion.   Further study can provide details on what markers, such as initial motor ability, predict outcomes using this device.  Some changes were found in the elbow and shoulder, which may be due stimulation to or other factors. Vibration can be widely conducted throughout the human body via bones and other tissues \cite{matsumoto1998dynamic, gurram1994vibration}.      

Future work should examine if these results are maintained, but the informal follow-ups that were accepted by the two participants suggest that improvements may be lasting.
Some participants had their stroke many years ago, and demonstrated change in measures.  
Participants used the device for over \textbf{160 hours each}, an intensity enabled by the wearable form factor and the passive stimulation method.  
In line with this result, a body of research has previously associated rehabilitation intensity (practice time) with improved outcomes \cite{teasell2005role, kwakkel2006impact, jette2005relation}.
Participants in the experimental condition reported new capabilities on the weekly worksheet that included helping to cook, cleaning their hobby equipment, donning winter gloves and holding their partner’s hand.  They also reported new tactile perception from the hand including sensing the vibrations, hyperextension during stretching, and the spray of water.  Three participants reported a greater sense of embodiment or ownership of the limb.  
Participants took advantage of the mobile nature of the device: reporting wearing the device to events such as church, lunch, and the movies.

\subsection*{Study limitations}
This investigation intends to establish the feasibility of wearable vibrotactile stimulation to improve diminished limb function.  Participants include various levels of disability, which provides initial data on who may be suited for this stimulation.    The Modified Ashworth Scale is a standard measure of tone and spasticity, but there are confounding factors for this measure.  These factors were controlled whenever possible including arm position, time of day, and rater.

\subsection*{Effects of the Control Condition}
Some change in measures may be expected when using the sham device.   
The sham device provided cutaneous sensory stimulation via the fabric of the glove; while the VTS experimental device provided additional cutaneous, and proprioceptive, stimulation via vibration.  
Furthermore, both conditions encouraged attention and engagement with the limb.  
 However, in contrast to the experimental group, the control group did not show significant changes.

\subsection*{Possible Mechanisms Behind Changes in Limb Function}
A wearable device can facilitate engagement with the disabled limb, which may help discourage maladaptive plastic changes from sensory deprivation and learned non-use.   Learned non-use \cite{grant2018somatosensory,liepert1998motor,wolf2006effect} is thought to be one of the reasons behind limited functional improvement of limbs after stroke: survivors learn to compensate and do not force themselves to re-learn the use of their limb.  
In addition, participants stretched open their affected hand to don and doff the device several times per day.  This stretching was expected to impact Modified Ashworth measures.  
Lesion location was not recorded in the study, but this information would provide interesting additional data if recorded in future work.

The control condition allowed us to examine the impacts of these mechanisms.  All participants interacted with a wearable device, but the experimental VTS group showed significantly different clinical measures after eight weeks.  This difference suggests that engagement and stretching may not be the only mechanisms to influence the participants.

Changes in tactile perception may be due central mechanisms.  Afferent input, transmitted by intact peripheral nervous pathways, may activate central nervous system regions.  This sensory input could impact central organization as is found in constraint-induced movement therapy after brain injury, or during normal sensorimotor skill acquisition 
\cite{schaechter2002motor, liepert1998motor, liepert2000treatment}.  

Vibration may help regulate electrophysiology associated with spasticity via afferent feedback. Reduced threshold of the stretch reflex has been implicated as one of the mechanisms behind symptoms of spasticity \cite{dietz2012spasticity, powers1988quantitative}. Supraspinal control usually regulates this reflex, but can be disrupted in events such as spinal cord injury or stroke \cite{dietz2012spasticity}. These reflexes are also mediated by afferent feedback produced during limb movement \cite{lynskey2008activity, sheean2002pathophysiology}.   Vibration provides similar feedback -- like many small muscle stretches -- activating cutaneous mechanoreceptors and proprioceptive afferents  \cite{burke1976responses, fallon2007vibration}.  
Afferent feedback then may induce reflex suppression and involuntary muscle contraction --  which may impact spasticity and are found during whole body vibration (WBV) and focal muscle/tendon vibration \cite{abercromby2007variation,delecluse2003strength,marconi2011long,ness2009effect,noma2009anti}.   
Presynaptic inhibition from afferent discharge is cited as a possible mechanism underlying reflex suppression during vibration \cite{rittweger2010vibration}.  Continuous passive motion is another treatment for spasticity, but removal of proprioceptive afferents was shown to prevent normalization \cite{lynskey2008activity, ollivier2010proprioceptive} suggesting that sensory feedback may underlie this method.   
Investigation of these factors is beyond the scope of this work, but the promising results warrant further study.

 Improved voluntary range of motion may be unlocked when spasticity and tone decreases.  
Another hypothesis for changes in voluntary motion is that such sensory stimulation provides excitatory feedback and coactivation of motor systems, and helps restore somatosensation useful in motor function \cite{cordo2009assisted,doyle2010interventions,platz2004impairment,raghavan2007nature}. This hypothesis is supported by work in sensory stimulation for motor learning and performance \cite{rosenkranz2003differential}, and motor rehabilitation \cite{cordo2013treatment,forner2008changes, golaszewski2012modulation, kaelin2002modulation}.

\section*{Conclusions}
A controlled, randomized trial of 16 participants evaluated the feasibility of a wearable vibrotactile stimulation method to reduce upper limb disability in chronic stroke.
All users were assigned to wear a computerized glove on their affected hand for three hours per day.  Users in the sham control group received no stimulation and those in the experimental condition received vibrotactile stimulation from the glove.

The wireless, wearable device was used during daily life, not in a clinical setting.
Participants who received vibrotactile stimulation demonstrated a significant change in measures of tactile perception, voluntary motion, and spasticity after eight weeks.
Some participants reported increase in protective sensation, sense of embodiment, and return to activities of daily living such as cleaning, cooking and writing using their disabled hand.

\textbf{\section*{Declarations}}
\vspace*{-20pt}
\begin{backmatter}
\textbf{ \subsection*{Ethical Approval and Consent to participate}}
 This study was approved and overseen by the Office of Research Integrity's IRB board for the Georgia Institute of Technology.    
All participants were screened using the Mini Mental State Exam (MMSE) and provided written consent before beginning the study.
 
\textbf{ \subsection*{Consent for publication}}
 Not Applicable. 
 
\textbf{ \subsection*{Availability of supporting data}}
 Not Applicable.
 
\textbf{ \subsection*{Competing interests}}
 The authors declare that they have no competing interests.
 
\textbf{  \subsection*{Funding}}
This research was supported, in part, by the National Science Foundation (NSF) Graduate Research Fellowship program, a grant from the Georgia Tech Graphics, Visualization and Usability (GVU) consortium, and a Microsoft Research PhD Fellowship. 

\textbf{ \subsection*{Authors' contributions}}
 All authors contributed to the study design and methodology. CES designed and fabricated the devices used in the research, recruited participants, and interfaced with clinicians. All authors drafted, edited, and approved the final manuscript.
 
\textbf{ \subsection*{Acknowledgements}}
 We would like to acknowledge and thank the other advisors to the project: Edelle Field-Fote (Shepherd Center/Emory University/Georgia Tech), Sarah Callahan (Shepherd Center) and Samir Belagaje (Emory University/Grady Hospital Stroke and Neuroscience Center).  
Thanks also to the clinicians of PT Solutions Midtown Atlanta and Physio Clinic Midtown Atlanta for taking measurements in this study.

 \hspace{1pt}
  
  \end{backmatter}

\bibliographystyle{bmc-mathphys} 
\bibliography{bmc_article}      

\end{document}